\pdfoutput=1
\documentclass[preprint,nonatbib,10pt]{elsarticle}

\usepackage[a4paper,top=2cm,bottom=2cm,left=3cm,right=3cm,marginparwidth=1.75cm]{geometry}

\usepackage[english]{babel} 
\usepackage{multicol} 

\usepackage[utf8]{inputenc}
\usepackage[T1]{fontenc}

\usepackage{lmodern} 
\usepackage[defaultlines=3,all]{nowidow} 

\usepackage{graphicx} 
\graphicspath{ {src/img/} }
\usepackage{tikz}
\usepackage{pgfplots}
\pgfplotsset{compat=1.17}
\usepackage{float} 
\usepackage{subcaption} 

\usepackage{amsmath}
\usepackage{amsfonts}
\usepackage{amssymb}
\usepackage{mathtools}
\usepackage{cancel} 
\usepackage{siunitx} 
\makeatletter
\let\c@author\relax
\makeatother

\usepackage[backend=biber,style=numeric,sorting=none]{biblatex}
\usepackage{csquotes}

\usepackage{lineno}

\newcommand{\DocumentTitle}{Generalized Treatment of Energy Accommodation in Gas-Surface Interactions for Satellite Aerodynamics Applications}
\newcommand{\DocumentAuthor}{Friedrich Tuttas, Constantin Traub, Marcel Pfeiffer, Walter Fichter}
\newcommand{\DocumentType}{Technical Note}

\usepackage[
	pdfauthor={\DocumentAuthor},
	pdftitle={\DocumentTitle},
	pdfsubject={\DocumentType},
	pdfkeywords={},
	pdfproducer={},
	pdfcreator={}
	]{hyperref} 

\DeclareMathOperator{\erf}{erf}
\DeclareMathOperator{\erfc}{erfc}

\nopreprintlinetrue

\begin{document}
\begin{frontmatter}
    \title{\DocumentTitle}

    \author[ifr]{Friedrich Tuttas\corref{cor1}}
    \author[irs]{Constantin Traub}
    \author[irs]{Marcel Pfeiffer}
    \author[ifr]{Walter Fichter}
    \cortext[cor1]{Corresponding author. Email address: \href{mailto:friedrich.tuttas@ifr.uni-stuttgart.de}{friedrich.tuttas@ifr.uni-stuttgart.de}}
    \affiliation[ifr]{organization={University of Stuttgart, Institute of Flight Mechanics and Controls},
        addressline={Pfaffenwaldring~27},
        postcode={70569},
        postcodesep={},
        city={Stuttgart},
        country={Germany}}
    \affiliation[irs]{organization={University of Stuttgart, Institute of Space Systems},
        addressline={Pfaffenwaldring 29},
        postcode={70569},
        postcodesep={},
        city={Stuttgart},
        country={Germany}}

    \begin{abstract}
        In the context of satellite aerodynamics in the Very-Low-Earth-Orbit (VLEO) regime, accurate modeling of gas-surface interactions (GSI) is crucial for determining aerodynamic forces and torques.
        Common models such as Sentman's assume that gas particles are reflected diffusely from a surface, which leads to the incorporation of energy accommodation into the model.
        This technical note discusses the limitations of existing approaches for handling energy accommodation and provides a generalized treatment thereof that is valid for any molecular speed ratio.
        A new general expression for the temperature ratio of reflected to impinging particles is derived, which, when used in a GSI model, retains its validity even in hypothermal flows.
        Additionally, a simplified hyperthermal approximation is presented, proven to be an asymptote of the general expression, and shown to be an improvement upon existing approximations by comparison for a realistic VLEO scenario.
        The results contribute to a better understanding and modeling of GSI, potentially benefiting scientific investigations and operational applications in satellite aerodynamics.
    \end{abstract}
\end{frontmatter}
\section{Introduction}
In the Very-Low-Earth-Orbit (VLEO) regime, the residual atmosphere is dense enough for the atmospheric drag force to represent the dominant perturbing effect acting on satellites.
At the same time, the density is still low enough for the flow to be considered a free molecular flow so that the interaction of incoming particles with reflected ones is negligible.
Consequently, it suffices to consider the momentum exchange of the particles with the satellite's external surfaces to determine the aerodynamic forces and torques.
To describe these gas-surface interactions (GSI), adequate models are required.
Since the underlying phenomena of this interaction are very complex, advances in modeling can, for example, improve the accuracy of scientific investigations like the estimation of thermospheric density and wind from satellite data \cite{doornbos_thermospheric_2012} or enable operational advantages such as aerodynamics-based, propellant-free attitude and orbit control of satellites \cite{livadiotti_application_2021}\cite{traub_differential_2023}. 

In terms of a suitable GSI model for satellite aerodynamics, Sentman's \cite{sentman_free_1961} is frequently employed and even described as the de facto standard for the computation of aerodynamic coefficients of satellites at low altitudes \cite{llop_spacecraft_2014}.
It assumes that the gas particles are reflected diffusely from a surface, simplifying the treatment of the momentum exchange by fully describing it through the energy exchange of these particles with the surface, which in turn depends on their temperature.

While Sentman already mentioned the relevance of a so-called energy accommodation coefficient~$\alpha_\mathrm{E}$, it was \textcite{moe_simultaneous_2004} who related it to the temperature of the reflected particles and thus introduced it as an input to Sentman's model.
In 2009, \textcite{koppenwallner_energy_2009} proposed a slightly different expression, which since then has been widely applied.
As it will be shown in a later stage of this note, however, the approximation can be used with small errors only for hyperthermal flows so that a more general consideration is still pending.

This note aims to contribute to the research field by providing a more generalized treatment of energy accommodation which is valid for any molecular speed ratio and retains the general validity of the GSI model even in hypothermal flows.
Furthermore, an approximation for hyperthermal flows is derived which, unlike the one presented in Ref.~\cite{koppenwallner_energy_2009}, is shown to be an asymptote of the general expression for the temperature ratio and therefore converges to it.
The contributions will help to enhance the understanding and modeling of gas-surface-interactions in VLEO and, as a consequence, scientific and operational efforts based on it could benefit in the long term. 

The note is structured as follows: in Section~\ref{sec:Background}, the theoretical background of the relation between GSI models, the temperature of reflected particles and energy accommodation is discussed.
Section~\ref{sec:TemperatureRatio} presents the derivation of a general expression for the temperature ratio of the reflected and the impinging particles, which represents the key contribution of this note.
In Section~\ref{sec:Approximation}, a simple asymptotic approximation for hyperthermal flows is derived from this general expression and subsequently compared to the existing approximation by \textcite{koppenwallner_energy_2009} in order to demonstrate the improvements.

\section{Theoretical Background}
\label{sec:Background}

\subsection{Gas Temperatures in Gas-Surface-Interaction Models}
\label{ssec:Relation}
A number of gas-surface-interaction models predict the forces acting on a structure in a free-molecular-flow environment based on diffuse reflection of all or at least some of the impinging gas particles from the structure's surface (e.g. Sentman~\cite{sentman_free_1961}).
This is based on the assumption that those particles lose all memory of the incident direction and can thus be modelled as being emitted from a fictitious gas in thermal equilibrium (i.e. with a Maxwellian velocity distribution).
Under these assumptions, the vector of the average reflected velocity is perpendicular to the surface.
Since its direction is known, the determination of this vector is simplified to finding its magnitude which can be accomplished by considering the average translational energy of the particles after the collision.
For a gas in thermal equilibrium, this energy is related to the temperature of the gas.

For this reason, the temperature of the reflected particles must be known in order to calculate the forces that the gas exerts on the surface.

\subsection{Energy Accommodation}
\label{ssec:EnergyAccommodation}
The average energy of a particle after a collision with a surface can be determined using the energy accommodation coefficient $\alpha_\mathrm{E}$.
It is defined as \cite{goodman_thermal_1974}:
\begin{align}
    \alpha_\mathrm{E} = \frac{\bar E_\mathrm{i}-\bar E_\mathrm{r}}{\bar E_\mathrm{i} - \bar E_\mathrm{w}} \label{eq:alphaE}
\end{align}
where $\bar E_\mathrm{i}$ is the average energy of the impinging particles, $\bar E_\mathrm{r}$ is the average energy of particles being emitted through the surface element from a fictitious gas in thermal equilibrium with a temperature $T_\mathrm{r}$, and $\bar E_\mathrm{w}$ is the same expression as $\bar E_\mathrm{r}$ but with the temperature of the surface $T_\mathrm{w}$.

When using this equation, it is important to carefully select the appropriate energy accommodation coefficient.
\textcite{goodman_thermal_1974} mentions that a distinction has to be made between equilibrium and non-equilibrium energy accommodation coefficients.
The former is only defined for stationary gases in thermal equilibrium with the surface while the latter would also be valid for gases in motion.
Furthermore, gas molecules generally possess internal energies, most notably due to their rotational and vibrational degrees of freedom, in addition to translational energy.
According to \textcite{goodman_thermal_1974}, an energy accommodation coefficient could be defined for each of these terms or any combination thereof.
GSI models, however, focus on momentum exchange and thus only consider the translational energy of the gas particles.
As a consequence, the proper coefficient to be used for determining the forces acting on a structure in a GSI model would be a \emph{non-equilibrium translational energy accommodation coefficient}.

Obtaining this specific coefficient through experiments is difficult.
However, since the atmosphere in the VLEO regime is dominated by atomic oxygen~\cite{crisp_experimental_2022}, it can be modelled as a monatomic gas.
As such, its particles do not possess rotational or vibrational degrees of freedom.
If this assumption also turned out to be valid during and after the collision with the surface, it would be a natural simplification to use the accommodation coefficient for the total energy instead of for the translational portion.

In the literature, $\alpha_\mathrm{E}$ is typically assumed to be in the range of about 0.9 to 1.0 for VLEO applications \cite{moe_gassurface_2005} indicating a relatively high degree of energy accommodation.

\subsection{The Temperature of the Reflected Particles}
\label{ssec:TempReflect}
Under the assumption that the correct $\alpha_\mathrm{E}$ is known and with the above definitions, finding the temperature of the reflected particles is a matter of determining expressions that relate the energies to the inflow conditions as well as to $T_\mathrm{r}$ and $T_\mathrm{w}$ and inserting these into Eq.~\ref{eq:alphaE}. 
\mbox{\textcite{moe_simultaneous_2004}} use the average translational energy of monatomic gas particles in thermal equilibrium within a given volume for both $\bar E_\mathrm{r}$ and~$\bar E_\mathrm{w}$:
\begin{align}
    \bar E_\textnormal{r/w} &= \frac{3}{2} k T_\textnormal{r/w}
\end{align}
with $k$ being the Boltzmann constant.
This, however, is invalid as it is not equal to the average translational energy of the particles that are emitted through a surface element. 
\mbox{\textcite{koppenwallner_energy_2009}} therefore correctly applied:
\begin{align}
    \bar E_\textnormal{r/w} &= 2 k T_\textnormal{r/w}. \label{eq:ErEw}
\end{align}
Within the GSI models, the temperature $T_\mathrm{r}$ appears in the form of the temperature ratio $T_\mathrm{r} / T_\mathrm{i}$, with $T_\mathrm{i}$ being the temperature of the impinging gas particles.
Inserting Eq.~\ref{eq:ErEw} into Eq.~\ref{eq:alphaE} and rearranging the result yields the following expression for the desired ratio:
\begin{align}
    \frac{T_\mathrm{r}}{T_\mathrm{i}} = \frac{\bar E_\mathrm{i}}{2 k T_\mathrm{i}} \left( 1 - \alpha_\mathrm{E} \right) + \alpha_\mathrm{E} \frac{T_\mathrm{w}}{T_\mathrm{i}}. \label{eq:TrTi_Ti}
\end{align}
Using the definitions of the molecular speed ratio $s$
\begin{equation}
	s = \frac{V_\mathrm{i} }{c_\mathrm{m}} 
\end{equation}
where $V_\mathrm{i}$ is the magnitude of the average velocity of the impinging particles and $c_\mathrm{m}$ the most probable thermal velocity of the impinging gas particles, which is defined as
\begin{equation}
 c_\mathrm{m} = \sqrt{  \frac{2 k T_\mathrm{i}}{m}},
\end{equation}
a relation between $T_\mathrm{i}$ and $V_\mathrm{i}$ can be obtained:
\begin{align}
    T_\mathrm{i} = \frac{m V_\mathrm{i}^2}{2 k s^2} \label{eq:TiVi}.
\end{align}
Here, $m$ is the mass of a gas particle.
This can be used to replace the dependency of the right side of Eq.~\ref{eq:TrTi_Ti} on $T_\mathrm{i}$ by a dependency on $s$ and $V_\mathrm{i}$:
\begin{align}
    \frac{T_\mathrm{r}}{T_\mathrm{i}} = \frac{s^2}{2} \left[ \frac{2 \bar E_\mathrm{i}}{m V_\mathrm{i}^2} \left( 1 - \alpha_\mathrm{E} \right) + \alpha_\mathrm{E} \frac{4 k T_\mathrm{w}}{m V_\mathrm{i}^2} \right]. \label{eq:TrTi_Vi}
\end{align}
In summary, Eq.~\ref{eq:TrTi_Vi} is an expression for the temperature ratio that was derived by inserting the above definitions into that of the energy accommodation coefficient in Eq.~\ref{eq:alphaE}.
The remaining task is to determine the average energy of the impinging particles $\bar E_\mathrm{i}$ and to insert it into Eq.~\ref{eq:TrTi_Vi} to obtain the desired general expression for the temperature ratio.

\section{Generalized Temperature Ratio}
\label{sec:TemperatureRatio}

\subsection{Average Energy of Impinging Particles}
\label{ssec:EnergyImpinging}
For the average energy of the impinging particles $\bar E_\mathrm{i}$, both \textcite{moe_simultaneous_2004} as well as \textcite{koppenwallner_energy_2009} use the following relation:
\begin{align}
    \bar E_\mathrm{i} = \frac{1}{2} m V_\mathrm{i}^2. \label{eq:Ei_hyp}
\end{align}
This, however, represents a hyperthermal approximation that is only valid for high molecular speed ratios as will be explained in the following.

In general, the average energy of the impinging particles can be determined as the ratio of the energy flux $\varepsilon_\mathrm{i}$ and the particle flux $\nu_\mathrm{i}$ through a surface element.
Both these quantities were derived by \textcite{bird_molecular_1994} (Eq.~4.27 and 4.22) as follows:
\begin{align}
	\varepsilon_\mathrm{i} &= \frac{m n c_\mathrm{m}^3}{4} \left[\frac{1}{\sqrt{\pi}}(s^2 + 2) \exp (-s^2 \cos^2 \delta) +s \cos\delta \left(s^2 + \frac{5}{2}\right) (1+\erf(s\cos\delta)) \right]\footnotemark,\\
    \nu_\mathrm{i} &= \frac{n c_\mathrm{m}}{2} \left[\frac{1}{\sqrt{\pi}} \exp(-s^2\cos^2\delta) + s\cos\delta (1+ \erf(s\cos\delta)) \right].
\end{align}
\footnotetext{Eq.~4.27 of Ref.~\cite{bird_molecular_1994} erroneously features an exponent of 2 for $c_\mathrm{m}$. This note uses the correct exponent of 3.}
Here, $n$ is the number density of the gas, $\delta$ the angle between the average velocity vector of the impinging particles and the surface normal, and $\erf$ the error function.
The average energy of the impinging particles is consequently:
\begin{align}
    \bar E_\mathrm{i} &= \frac{\varepsilon_\mathrm{i}}{\nu_\mathrm{i}} \\
    &= \frac{m c_\mathrm{m}^{2}}{2} \left[ 2 + s^2 + \frac{1}{2} \frac{ s \cos\delta \left( 1 + \erf \left( s \cos\delta \right) \right) }{\frac{1}{\sqrt{\pi}} \exp \left( - s^2 \cos^2\delta \right) + s \cos\delta \left( 1 + \erf \left( s \cos\delta \right) \right)} \right] \\
    &= k T_\mathrm{i} \left[ 2 + s^2 + \frac{1}{2} \frac{ s \cos\delta \left( 1 + \erf \left( s \cos\delta \right) \right) }{\frac{1}{\sqrt{\pi}} \exp \left( - s^2 \cos^2\delta \right) + s \cos\delta \left( 1 + \erf \left( s \cos\delta \right) \right)} \right]. \label{eq:EiTi}
\end{align}

For comparison with Eq.~\ref{eq:Ei_hyp}, Eq.~\ref{eq:EiTi} can also be transformed so that its explicit dependency on $T_\mathrm{i}$ is replaced by an explicit dependency on $V_\mathrm{i}$ using Eq.~\ref{eq:TiVi}:
\begin{align}
    \bar E_\mathrm{i}
    &= \frac{m V_\mathrm{i}^{2}}{2} \left[ \frac{2}{ s^2} + 1 + \frac{1}{2 s^2} \frac{ s \cos\delta \left( 1 + \erf \left( s \cos\delta \right) \right) }{\frac{1}{\sqrt{\pi}} \exp \left( - s^2 \cos^2\delta \right) + s \cos\delta \left( 1 + \erf \left( s \cos\delta \right) \right)} \right]. \label{eq:EiVi_1-erf}
\end{align}
This is a general expression for the average energy of particles of a gas in thermal equilibrium and with a velocity $V_\mathrm{i}$ that collide with a surface at an angle of $\delta$ with the surface normal.

It should be noted that a term like $1 - \erf\, x$ can exhibit numerical problems for large values of $x$ which results in a loss of precision.
This occurs in Eq.~\ref{eq:EiVi_1-erf} for large values of $s$ and when $\cos\delta < 0$.
An implementation of the complementary error function $\erfc x = 1 - \erf x$ that retains a high accuracy for large values of $x$ exists for many programming languages.
Using this function, Eq.~\ref{eq:EiVi_1-erf} can be rewritten into a form that is more suitable for computational implementations:
\begin{align}
    \bar E_\mathrm{i} &= \frac{m V_\mathrm{i}^{2}}{2} \left[ \frac{2}{ s^2} + 1 + \frac{1}{2 s^2} \frac{ s \cos\delta \erfc\left( - s \cos\delta \right) }{\frac{1}{\sqrt{\pi}} \exp \left( - s^2 \cos^2\delta \right) + s \cos\delta \erfc\left( - s \cos\delta \right)} \right]. \label{eq:EiVi}
\end{align}

In the hyperthermal case, the average velocity $V_\mathrm{i}$ is much larger than the most probable thermal velocity $c_\mathrm{m}$  so that large values for the molecular speed ratio $s$ result.
Investigating the limit of the average energy as $s$ approaches infinity via Eq.~\ref{eq:EiVi} shows that the expression in brackets tends to one for surfaces that face the flow (i.e. $\cos\delta > 0$).
Thus, the overall limit of the average energy is equal to Eq.~\ref{eq:Ei_hyp} which verifies the statement that Eq.~\ref{eq:Ei_hyp} represents a hyperthermal approximation.

\subsection{General Expression for the Temperature Ratio}
\label{ssec:TemperatureRatio}
In order to arrive at a result equivalent to Eqs. 26a and 42 of Ref.~\cite{koppenwallner_energy_2009}, Eq.~\ref{eq:Ei_hyp} needs to be inserted into Eq.~\ref{eq:TrTi_Vi} which yields:
\begin{align}
    \frac{T_\mathrm{r}}{T_\mathrm{i}} &= \frac{s^2}{2} \left[ 1 + \alpha_\mathrm{E} \left( \frac{4 k T_\mathrm{w}}{m V_\mathrm{i}^2} - 1 \right) \right]. \label{eq:TrTi_kop}
\end{align}
Since Eq.~\ref{eq:Ei_hyp} was shown to be a hyperthermal approximation, the resulting expression for the temperature ratio in Eq.~\ref{eq:TrTi_kop} also represents a hyperthermal approximation that is only valid as $s$ approaches infinity.

The general expression for the temperature ratio, which is valid for any molecular speed ratio, can be obtained by inserting Eq.~\ref{eq:EiVi} (instead of Eq.~\ref{eq:Ei_hyp}) into Eq.~\ref{eq:TrTi_Vi}.
The resulting expression, which represents a key contribution of this note, is:
\begin{alignat}{2}
    \frac{T_\mathrm{r}}{T_\mathrm{i}} &=\, &&\frac{s^2}{2} \left[ \alpha_\mathrm{E} \frac{4 k T_\mathrm{w}}{m V_\mathrm{i}^2} + \left( 1 - \alpha_\mathrm{E} \right) \left( \frac{2}{ s^2} + 1 + \frac{1}{2 s^2} \frac{ s \cos\delta \erfc\left( - s \cos\delta \right) }{\frac{1}{\sqrt{\pi}} \exp \left( - s^2 \cos^2\delta \right) + s \cos\delta \erfc\left( - s \cos\delta \right)} \right) \right] \\
    &=\, && \alpha_\mathrm{E} \frac{2 k T_\mathrm{w}}{m V_\mathrm{i}^2} s^2 + \left( 1 - \alpha_\mathrm{E} \right) \left( 1 + \frac{s^2}{2} + \frac{1}{4} \frac{ s \cos\delta \erfc\left( - s \cos\delta \right) }{\frac{1}{\sqrt{\pi}} \exp \left( - s^2 \cos^2\delta \right) + s \cos\delta \erfc\left( - s \cos\delta \right)} \right)
    \label{eq:TrTi}
\end{alignat}

\section{Hyperthermal Approximation}
\label{sec:Approximation}

\subsection{Derivation of an Asymptotic Hyperthermal Approximation}
\label{ssec:Derivation}
The general expression for the temperature ratio presented in Eq.~\ref{eq:TrTi} can be used to obtain a simple hyperthermal approximation of the temperature ratio which represents an improvement to the approximation from Ref.~\cite{koppenwallner_energy_2009}. 
Contrary to the previous procedure, the limit as $s$ approaches infinity is determined directly for the general temperature ratio in Eq.~\ref{eq:TrTi} instead of for the average energy in Eq.~\ref{eq:EiVi}:
\begin{alignat}{3}
	\lim_{s\to\infty} \frac{T_\mathrm{r}}{T_\mathrm{i}}
	&= \, &&\alpha_\mathrm{E} \frac{4 k T_\mathrm{w}}{m V_\mathrm{i}^2} \left( \lim_{s\to\infty} \frac{s^2}{2} \right) \nonumber\\
    &&&+ \left( 1 - \alpha_\mathrm{E} \right) \left[ 1 + \left( \lim_{s\to\infty} \frac{s^2}{2} \right) + \frac{1}{4} \left( \lim_{s\to\infty} \frac{ s \cos\delta \erfc\left( - s \cos\delta \right) }{\frac{1}{\sqrt{\pi}} \exp \left( - s^2 \cos^2\delta \right) + s \cos\delta \erfc\left( - s \cos\delta \right)} \right) \right].
\end{alignat}

Regarding the last limit term in brackets, a distinction has to be made between the cases where $\cos\delta$ is positive and where it is negative.
The first case is the more relevant one since $\cos\delta > 0$ means that the surface is facing the flow.
It is also easier to find an asymptote for this case since the last term will tend to 1 as $s$ approaches infinity eliminating the dependency on the angle $\delta$:
\begin{align}
	\lim_{s\to\infty} \frac{T_\mathrm{r}}{T_\mathrm{i}} &= \left[ 1 + \alpha_\mathrm{E} \left( \frac{4 k T_\mathrm{w}}{m V_\mathrm{i}^2} - 1 \right) \right] \left( \lim_{s\to\infty} \frac{s^2}{2} \right) + \frac{5}{4} \left( 1 - \alpha_\mathrm{E} \right).
\end{align}
According to this, the following equation is an asymptote for the temperature ratio as the molecular speed ratio tends to infinity and thus can serve as a hyperthermal approximation for flow-facing surfaces:
\begin{align}
	\frac{T_\mathrm{r}}{T_\mathrm{i}} \approx \frac{s^2}{2} \left[ 1 + \alpha_\mathrm{E} \left( \frac{4 k T_\mathrm{w}}{m V_\mathrm{i}^2} - 1 \right) \right] + \frac{5}{4} \left( 1 - \alpha_\mathrm{E} \right). \label{eq:TrTi_hyp}
\end{align}
While both approximations (Eq.~\ref{eq:TrTi_hyp} and the result of Ref.~\cite{koppenwallner_energy_2009} in Eq.~\ref{eq:TrTi_kop}) share the general solution's property of tending to infinity with growing $s$, a comparison shows a constant offset of ${5 \left( 1 - \alpha_\mathrm{E} \right) / 4}$ between them.
Therefore, Eq.~\ref{eq:TrTi_kop} cannot be an asymptote for the general solution of the temperature ratio.

Finding an approximation like Eq.~\ref{eq:TrTi_hyp} for the rear-facing surfaces is more difficult since the last term in brackets does not simply tend to a constant value now.
For completeness' sake, an asymptotic approximation shall also be mentioned for the cases where $\cos\delta < 0$ while its derivation is omitted:
\begin{align}
	\frac{T_\mathrm{r}}{T_\mathrm{i}} \approx \frac{s^2}{2} \left[ 1 - \cos^2\delta + \alpha_\mathrm{E} \left( \frac{4 k T_\mathrm{w}}{m V_\mathrm{i}^2} - \left( 1 - \cos^2\delta \right) \right) \right] + \frac{1}{2} \left( 1 - \alpha_\mathrm{E} \right). \label{eq:TrTi_hyp_rear}
\end{align}
A proof that this equation is an asymptote of the general solution is given in~\ref{app:Asymptoticity}.
As can be seen, the approximation for these surfaces still depends on the angle $\delta$ unlike the one for the flow-facing surfaces.
Since the hyperthermal approximation of Sentman's model disregards the rear-facing surfaces altogether, Eq.~\ref{eq:TrTi_hyp_rear} is irrelevant for the calculation of aerodynamic forces in practice.

\subsection{Asymptoticity of Hyperthermal Approximations}
\label{ssec:ApproximationDiscussion}
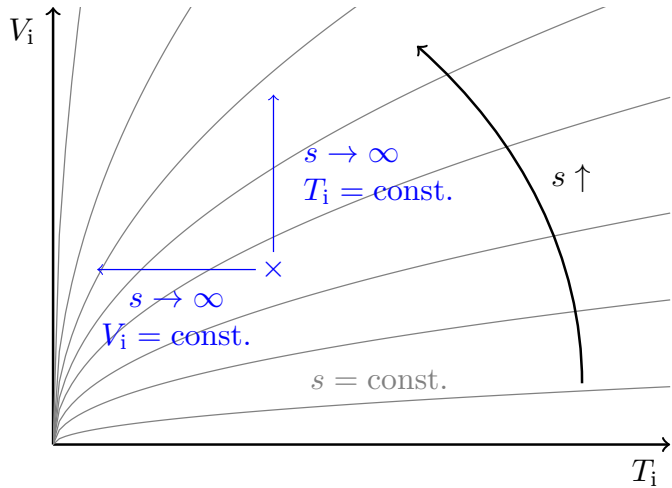
\begin{figure}
    \centering
    \resizebox{0.6\textwidth}{!}{\begin{tikzpicture}
    \begin{axis}[
        axis lines = middle,
        xmin=0, xmax=7,
        ymin=0, ymax=5,
        samples=100,
        domain=0:7,
        axis line style={->, thick},
        xtick=\empty,  
        ytick=\empty,  
        xlabel={$T_\mathrm{i}$},
        ylabel={$V_\mathrm{i}$},
        xlabel style={at={(axis description cs:1,-0.01)},anchor=north east},
        ylabel style={at={(axis description cs:-0.01,1)},anchor=north east},
        disabledatascaling,
        axis equal image,
    ]

    \addplot[gray] {0.25*sqrt(x)};
    \addplot[gray] {0.625*sqrt(x)};
    \addplot[gray] {sqrt(x)};
    \addplot[gray] {1.5*sqrt(x)};
    \addplot[gray] {2*sqrt(x)};
    \addplot[gray] {2.7*sqrt(x)};
    \addplot[gray] {4*sqrt(x)};
    \addplot[gray] {9*sqrt(x)};

    \node[gray] at (axis cs:3.7,0.75) {$s = \textnormal{const.}$};

    \draw[->, thick] 
        (axis cs:6,0.7) arc[start angle=0, end angle=40, x radius=8, y radius=6] node[midway, above right] {$s\uparrow$};
        
    \node[blue] at (axis cs:2.5,2) {$\times$};

    \draw[->, blue] (axis cs:2.3,2) -- (axis cs:0.5,2)
        node[midway, below, blue] 
        {\begin{tabular}{c} $s\to\infty$\\ $V_\mathrm{i} = \textnormal{const.}$\end{tabular}};

    \draw[->, blue] (axis cs:2.5,2.2) -- (axis cs:2.5,4)
        node[midway, right, blue]
        {\begin{tabular}{l} $s\to\infty$\\ $T_\mathrm{i} = \textnormal{const.}$\end{tabular}};

    \end{axis}
\end{tikzpicture}}
    \caption{Different Hyperthermal Approximations.}
    \label{fig:difhypappr}
\end{figure}

The dependency on $T_\mathrm{i}$ and $V_\mathrm{i}$ of neither the average energy nor the temperature ratio is completely contained within the molecular speed ratio $s$.
Even though $s$ was introduced in both Eqs.~\ref{eq:EiVi} and \ref{eq:TrTi}, they still explicitly depend on $V_\mathrm{i}$.
Using Eq.~\ref{eq:TiVi}, the dependency on $V_\mathrm{i}$ can be replaced by a dependency on $T_\mathrm{i}$.
While this has no effect on the validity of the general equations, it might affect the hyperthermal approximations because in one case $V_\mathrm{i}$ and in the other case $T_\mathrm{i}$ is kept constant while $s$ approaches infinity.
This is shown in Fig.~\ref{fig:difhypappr} where multiple curves of constant $s$ are plotted in a $T_\mathrm{i}$-$V_\mathrm{i}$-plane.
From a given point (marked by a blue cross), the limit as $s$ approaches infinity can be evaluated by moving along two different paths: 1.) a path on which $V_\mathrm{i}$ is kept constant and 2.) a path on which $T_\mathrm{i}$ is kept constant.
For quantities that depend on $T_\mathrm{i}$ and $V_\mathrm{i}$ in this manner, there may be asymptotes for one of these directions that cannot simply be transformed using Eq.~\ref{eq:TiVi} and expected to be an asymptote for the other direction.
If, for example, an asymptote was found for the direction where $V_\mathrm{i}$ is kept constant, its expression will depend on this value of $V_\mathrm{i}$.
Using Eq.~\ref{eq:TiVi} to transform this expression into one that explicitly depends on $T_\mathrm{i}$ may not necessarily yield an asymptote for the other direction. 

This problem does not occur for the approximation derived in this note (Eq.~\ref{eq:TrTi_hyp}).
It is an asymptote for the direction where $V_\mathrm{i}$ is kept constant and can be transformed into an asymptote for the direction where $T_\mathrm{i}$ is kept constant using Eq.~\ref{eq:TiVi}.

\subsection{Assessment of the Approximation Accuracy}
\label{ssec:Comparison}
To assess the accuracy of the available approximations, their relative error $R$ can be considered.
It shall be defined as:
\begin{align}
    R = \left\lvert\frac{\tau - \tau_\textnormal{appr}}{\tau}\right\rvert.
\end{align}
Here, $\tau$ represents the temperature ratio as given by the general Eq.~\ref{eq:TrTi}, and $\tau_\textnormal{appr}$, the temperature ratio as given by one of the approximations (Eqs.~\ref{eq:TrTi_kop}, \ref{eq:TrTi_hyp}, or~\ref{eq:TrTi_hyp_rear}).

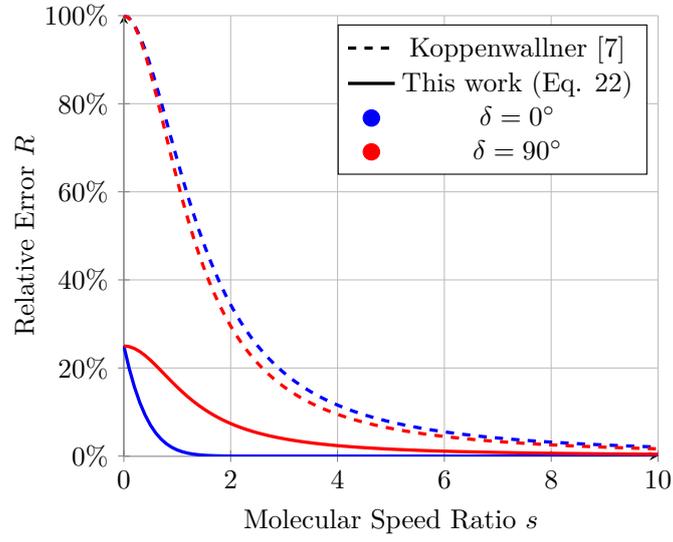
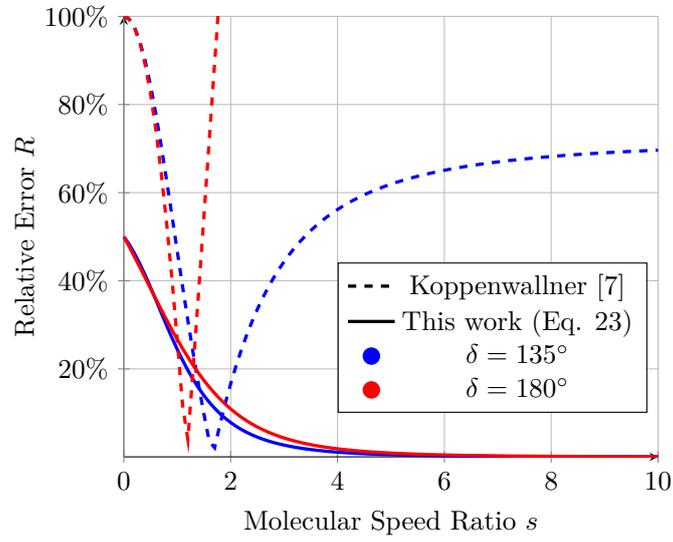
\begin{figure}
    \centering
    \begin{subfigure}{.6\linewidth}
        \centering
        \resizebox{\columnwidth}{!}{\begin{tikzpicture}
    \begin{axis}[
        axis lines = middle,
        extra x ticks = {0},
        extra y ticks= {0},
        xlabel={Molecular Speed Ratio $s$},
        ylabel={Relative Error $R$},
        yticklabel={\pgfmathparse{\tick*100}\pgfmathprintnumber{\pgfmathresult}\%},
        xlabel style={at={(axis description cs:0.5,-0.1)},anchor=north},
        ylabel style={at={(axis description cs:-0.15,0.5)}, rotate=90, anchor=south},
        grid
    ]        
        \addplot[blue, solid, very thick, forget plot] table [x=s, y=Rn, col sep=comma] {approximationComparison000.csv};
        \addplot[blue, dashed, very thick, forget plot] table [x=s, y=Rk, col sep=comma] {approximationComparison000.csv};
        \addplot[red, solid, very thick, forget plot] table [x=s, y=Rn, col sep=comma] {approximationComparison090.csv};
        \addplot[red, dashed, very thick, forget plot] table [x=s, y=Rk, col sep=comma] {approximationComparison090.csv};
        \addlegendimage{black, dashed, very thick}
        \addlegendentry{\textcite{koppenwallner_energy_2009}}
        \addlegendimage{black, solid, very thick}
        \addlegendentry{This work (Eq. \ref{eq:TrTi_hyp})}
        \addlegendimage{only marks, mark=*, mark size=3pt, blue}
        \addlegendentry{$\delta=\SI{0}{\degree}$}
        \addlegendimage{only marks, mark=*, mark size=3pt, red}
        \addlegendentry{$\delta=\SI{90}{\degree}$}
    \end{axis}
\end{tikzpicture}}
        \caption{Flow-facing surfaces.}
    \end{subfigure}
    \par\bigskip
    \begin{subfigure}{.6\linewidth}
        \centering
        \resizebox{\columnwidth}{!}{\begin{tikzpicture}
    \begin{axis}[
        axis lines = middle,
        extra x ticks = {0},
        extra y ticks= {0},
        xlabel={Molecular Speed Ratio $s$},
        ylabel={Relative Error $R$},
        yticklabel={\pgfmathparse{\tick*100}\pgfmathprintnumber{\pgfmathresult}\%},
        xlabel style={at={(axis description cs:0.5,-0.1)},anchor=north},
        ylabel style={at={(axis description cs:-0.15,0.5)}, rotate=90, anchor=south},
        grid,
        ymax = 1,
        legend style={at={(0.98,0.1)}, anchor=south east} 
    ]
        \addplot[blue, solid, very thick, forget plot] table [x=s, y=Rn_back, col sep=comma] {approximationComparison135.csv};
        \addplot[blue, dashed, very thick, forget plot] table [x=s, y=Rk, col sep=comma] {approximationComparison135.csv};
        \addplot[red, solid, very thick, forget plot] table [x=s, y=Rn_back, col sep=comma] {approximationComparison180.csv};
        \addplot[red, dashed, very thick, forget plot] table [x=s, y=Rk, col sep=comma] {approximationComparison180.csv};
        
        \addlegendimage{black, dashed, very thick}
        \addlegendentry{\textcite{koppenwallner_energy_2009}}
        \addlegendimage{black, solid, very thick}
        \addlegendentry{This work (Eq. \ref{eq:TrTi_hyp_rear})}
        \addlegendimage{only marks, mark=*, mark size=3pt, blue}
        \addlegendentry{$\delta=\SI{135}{\degree}$}
        \addlegendimage{only marks, mark=*, mark size=3pt, red}
        \addlegendentry{$\delta=\SI{180}{\degree}$}
    \end{axis}
\end{tikzpicture}}
        \caption{Rear-facing surfaces.}
    \end{subfigure}
    \caption{Relative error of approximations for increasing molecular speed ratios.} \label{fig:apprcomp}
\end{figure}

In Fig.~\ref{fig:apprcomp}, the relative error of all these approximations is plotted for increasing molecular speed ratios.
Since the general solution as well as the approximation for rear-facing surfaces depend on $\delta$, the relative error is evaluated for four different values (two for flow-facing and another two for rear-facing surfaces).
For the remaining parameters, a plausible set of values is chosen that could be encountered in a VLEO satellite mission.
These are listed in Tab.~\ref{tab:Parameter}.
\begin{table}
   \centering
   \caption{Parameters used for the assessment of the approximations.}
   \begin{tabular}{cc}
       \hline
       Parameter & Value \\
       \hline
       $T_\mathrm{w} $  & \SI{300}{\kelvin} \\
       $V_\mathrm{i} $  & \SI{7.8}{\kilo\meter\per\second} \\
       $\alpha_\mathrm{E} $  & 0.95 \\
       $m$ &  \SI{2.72e-26}{\kilo\gram} \\
       \hline
   \end{tabular}
   \label{tab:Parameter}
\end{table}

For the flow-facing surfaces, it is clearly visible that the relative error of the new approximation Eq.~\ref{eq:TrTi_hyp} derived in this note (labeled \enquote{This work}) has a significantly smaller value for $s = 0$ compared to that of Ref.~\cite{koppenwallner_energy_2009} and converges much quicker to zero.
Specifically for the scenario where $\delta = \SI{0}{\degree}$ (i.e. incoming flow is perpendicular to the surface), it has decreased below \SI{1.4}{\percent} for $s \geq 1$.
The approximation in Ref.~\cite{koppenwallner_energy_2009} (labeled \enquote{Koppenwallner}), on the other hand, shows a relative error of more than \SI{67}{\percent} at $s = 1$ which does not decrease below \SI{2}{\percent} until $s = 10$.

For rear-facing surfaces, the relative error of the respective approximation of Eq.~\ref{eq:TrTi_hyp_rear} starts with a higher value but still converges to zero while that of Ref.~\cite{koppenwallner_energy_2009} converges to a nonzero value.

\section{Conclusion and Outlook}
\label{sec:Conclusion}
Within this technical note, a generalized treatment of energy accommodation in GSI models is discussed resulting in a general expression of the temperature ratio of impinging and reflected particles being added to the state of knowledge.
It is shown to be valid for any molecular speed ratio and therefore retains the general validity of the GSI model even in hypothermal flows.
Subsequently, a simplified approximation for hyperthermal flows is derived which is shown to asymptotically converge to the general expression.

The results of this note are believed to enhance the understanding of incorporating energy accommodation into GSI models with (partial) diffuse reflection.
If the general expression for the temperature ratio (Eq.~\ref{eq:TrTi}) derived in this note is used in such a model, its general validity for any value of $s$ is retained.
This would extend the usability of those models to suborbital flights with hypothermal flows for which no accurate treatment of the energy accommodation exists in the literature. 
In hyperthermal regimes, an approximation of the temperature ratio for very high molecular speed ratios might suffice.
Here, the application of the one derived in this note (Eq.~\ref{eq:TrTi_hyp}) does not add any impractical complexity, as it only differs from the result in Ref.~\cite{koppenwallner_energy_2009} by a constant offset. 
Unlike the latter, the new expression of this note was shown to be an asymptote of the general expression and therefore converges to it with increasing $s$ yielding a lower relative error. 

Since the temperature ratio is only needed to determine the forces due to reflected particles and their velocity is generally much smaller than that of the impinging particles (which is of the order of the satellite's orbital velocity), the reflection plays a minor role in the force computation in practice.
Thus, the correction thereof presented within this note is only expected to have a small impact on the overall force computation.
It remains to be shown how much of a difference the usage of the general expression for the temperature ratio or the new approximation will make in the simulation of GSI models and subsequently in practical applications like the computation of thermospheric density data from satellite dynamics.

\section*{Acknowledgements}
This work is funded by the Deutsche Forschungsgemeinschaft project number 516238647 - SFB1667/1 (ATLAS - Advancing Technologies for Low-Altitude Satellites).
\printbibliography
\appendix
\section{Proof of the Asymptoticity of the Hyperthermal Approximation for Rear-Facing surfaces}
\label{app:Asymptoticity}
In order to prove that the approximation for rear-facing surfaces is an asymptote of the general expression of the temperature ratio, the difference of the two expressions must be shown to converge to zero.
If $\tau$ denotes the general expression of Eq.~\ref{eq:TrTi} and $\tau_\text{appr}$ the approximation of Eq.~\ref{eq:TrTi_hyp_rear}, their difference can be rewritten as:
\begin{alignat}{3}
    \tau - \tau_\text{appr} &= \,&&\alpha_\mathrm{E} \frac{2 k T_\mathrm{w}}{m V_\mathrm{i}^2} s^2 + \left( 1 - \alpha_\mathrm{E} \right) \left( 1 + \frac{s^2}{2} + \frac{1}{4} \frac{ s \cos\delta \erfc\left( - s \cos\delta \right) }{\frac{1}{\sqrt{\pi}} \exp \left( - s^2 \cos^2\delta \right) + s \cos\delta \erfc\left( - s \cos\delta \right)} \right) \nonumber\\
    &&&- \frac{s^2}{2} \left[ 1 - \cos^2\delta + \alpha_\mathrm{E} \left( \frac{4 k T_\mathrm{w}}{m V_\mathrm{i}^2} - \left( 1 - \cos^2\delta \right) \right) \right] - \frac{1}{2} \left( 1 - \alpha_\mathrm{E} \right) \\
    &= &&\frac{1}{4} \left( 1 - \alpha_\mathrm{E} \right) \left[ \frac{\frac{2}{\sqrt{\pi}} \left( 1+s^2 \cos^2\delta \right) \exp\left(-s^2\cos^2\delta\right) + \left( 3 + 2s^2 \cos^2\delta \right) s \cos\delta \erfc\left(-s \cos\delta\right)}{\frac{1}{\sqrt{\pi}} \exp\left(-s^2\cos^2\delta\right) + s \cos\delta \erfc\left(-s \cos\delta\right)} \right].
\end{alignat}
When $\cos\delta < 0$, both the enumerator and the denominator of the term in brackets tend to zero as $s$ approaches infinity.
L'Hôpital's rule can be applied to determine the limit of this term.
Repeated application of this rule leads to the following four terms (labeled $L_1$ through $L_4$), the last of which allows for a direct evaluation of the limit:
\begin{align}
    L_1 &= \frac{ \frac{6}{\sqrt{\pi}} s \cos^2\delta \exp\left( -s^2 \cos^2\delta \right) + 3 \left( 1 + 2 s^2 \cos^2\delta \right) \cos\delta \erfc\left( -s \cos\delta \right) }{ \cos\delta \erfc\left( -s \cos\delta \right) }, \\
    L_2 &= \frac{ \frac{12}{\sqrt{\pi}} \cos^2\delta \exp\left( -s^2 \cos^2\delta \right) + 12 s \cos^3\delta \erfc\left( -s \cos\delta \right) }{ \frac{2}{\sqrt{\pi}} \cos^2\delta \exp\left( -s^2 \cos^2\delta \right) }, \\
    L_3 &= \frac{ 12 \cos^3\delta \erfc\left( -s \cos\delta \right) }{ - \frac{4}{\sqrt{\pi}} s \cos^4\delta \exp\left( -s^2 \cos^2\delta \right) }, \\
    L_4 &= \frac{ \frac{24}{\sqrt{\pi}} \cos^4\delta \exp\left( -s^2 \cos^2\delta \right) }{ \frac{4}{\sqrt{\pi}} \cos^4\delta \left( 2s^2\cos^2\delta - 1 \right) \exp\left( -s^2 \cos^2\delta \right) } \\
    &= \frac{ 6 }{ 2 s^2 \cos^2\delta - 1 }.
\end{align}
The limit of $L_4$ as $s$ approaches infinity is equal to zero.
Therefore, the difference of the general expression and the approximation for rear-facing surfaces also tends to zero, which proves the asymptoticity of the approximation in Eq.~\ref{eq:TrTi_hyp_rear}.
\end{document}